\documentclass[aps,prb,reprint]{revtex4-1}
\usepackage{amsmath}
\usepackage{graphicx}
\usepackage{units}  
\usepackage{xspace}
\usepackage{subfigure}
\usepackage{hyperref}

\newcommand{\ve}{\varepsilon}
\newcommand{\mb}{\mathbf}
\newcommand{\tb}{\textbf}
\newcommand{\beq}{\begin{equation}}
\newcommand{\eeq}{\end{equation}}
\newcommand{\bea}{\begin{eqnarray}}
\newcommand{\eea}{\end{eqnarray}}

\usepackage[usenames]{color}

\begin{document}

\bibliographystyle{apsrev}
 
\title{Quantum Oscillations from Fermi Sea}

\date{\today}
\author{Hridis K. Pal}
   \email{hridis.pal@physics.gatech.edu}
\affiliation{LPS, CNRS UMR 8502, Univ. Paris-Sud, Univ. Paris-Saclay, 91405  Orsay Cedex, France}

\begin{abstract} 
Quantum oscillations are conventionally understood to arise from the Fermi level; hence, they are considered to be a proof of the existence of an underlying Fermi surface. In this article, we show that in certain situations quantum oscillations can also arise from inside the Fermi sea. We establish this analytically, supporting it with numerical calculations. Possible scenarios where such unusual behavior can occur are pointed out. In particular, in strongly particle-hole asymmetric insulators, models of which have been recently used in the context of the topological Kondo insulator SmB$_6$, we show that the oscillations arise from inside the filled band, and are not related to the gap.
\end{abstract}

\pacs{}
\maketitle 

\section{Introduction}

Quantum oscillations arise in metals when quantized energy levels in a magnetic field cross the Fermi level periodically as a function of the field, resulting in oscillations in physical observables. The phenomenon is ubiquitous: such oscillations appear in a variety of systems, ranging from simple band materials to strongly correlated systems, as long as the temperature is low enough and disorder is sufficiently weak \cite{sho}. As a result, measurement of quantum oscillations has become a standard experimental tool to study new materials. There are two key ingredients contributing to the phenomenon: quantized energy levels and the Fermi level---the latter provides a natural cutoff to the occupation of the energy levels at zero temperature. Because these oscillations arise from the Fermi level, observation of quantum oscillations is considered as a proof of the existence of an underlying Fermi surface. 

Recently, Tan et al. have reported \cite{tan} the observation of quantum oscillations in SmB$_6$, a Kondo insulator. Not unexpectedly, a large part of the effort to understand the experiment has been directed towards constructing a theory that yields a Fermi surface in the gap \cite{ert,bas,seb2,tho,den}. In parallel, a handful of other works \cite{kno, zha, pal} have considered the possibility of quantum oscillations in insulators without a Fermi surface. It has been shown that in insulators with narrow inverted bands such that the valence band (VB) edge forms a closed loop, oscillations can arise from this edge which now provides a cutoff to the occupation of states, thus playing a role similar to the Fermi level in metals \cite{zha,pal}. In light of such findings, the existence of a Fermi surface seems to be no longer a strict necessity for quantum oscillations. Instead, it is tempting to generalize that oscillations arise from the highest occupied energy state: in metals this is provided by the Fermi level, while in insulators this is provided by the edge of the VB. 

In this article, we show that the above generalization does not exhaust all possibilities: in certain situations, surprisingly, quantum oscillations can also arise from inside the band, i.e., inside the Fermi sea. In particular, in the kinds of insulators described above, if the insulator is strongly particle-hole asymmetric---as in the case of SmB$_6$---oscillations originate from inside the valence band and are not related to the gap. 

To understand qualitatively the origin, consider the grand canonical potential at $T=0$ (we set $\hbar=k_{B}=1$):
\beq
\Omega=D\sum_{E_n\le\mu}[E_n-\mu],
\label{omegadef}
\eeq
where $D\propto B$ is the degeneracy of each Landau level $E_n$, and $\mu$ is the Fermi level.
An equivalent way to write Eq.~(\ref{omegadef}) is
$\Omega=D\sum_{\mathcal{E}_n}[\mathcal{E}_n-\mu]$,
where
$\mathcal{E}_n$ is defined to be piecewise continuous: $\mathcal{E}_n=E_n$ for $E_n\le \mu$ and $\mathcal{E}_n=0$ for $E_n> \mu$, and the sum over the states is now unrestricted. Evidently the derivative $\mathcal{E}_n'\equiv\partial \mathcal{E}_n/\partial n$ is discontinuous at $\mathcal{E}=\mu$. Quantum oscillations are simply a manifestation of this discontinuity in the function $\mathcal{E}_n'$. Imagine now that the discontinuity is smoothened out on some scale $\zeta$. As long as $\zeta\ll\omega_c$, where $\omega_c$ is the typical Landau level spacing, although $\mathcal{E}_n'$ is no longer discontinuous, it still changes sharply. With change in field the Landau levels will still feel this abruptness and manifest as quantum oscillations (with reduced amplitude). Formulated this way, as far as quantum oscillations are concerned, the role of $\mu$ is simply to produce a feature in the function $\mathcal{E}_n$ such that its derivative changes sharply. Any such feature arising from some other source at a different energy, $E=\tilde{E}$, should similarly produce quantum oscillations, even if it is inside the Fermi sea. 

\section{Theory}

Let us now give the above argument a formal structure. To carry out the discrete sum in Eq.~(\ref{omegadef})
we use the Euler-Maclaurin formula:
\begin{eqnarray}
\sum_{r=0}^Rf(r)&=&\int_0^Rf(r)dr+\frac{1}{2}[f(R)+f(0)]\nonumber\\
&+&\frac{1}{12}[f'(R)-f'(0)]+\cdots,
\label{eulermac}
\end{eqnarray}
where $r$ on the right hand side is treated as a continuous variable. The formula simply states that the integral of a function in some interval can be approximated by a sum over discrete values of the function within that interval. The approximation works well as long as the number of discrete points summed over is large so that the function does not change significantly between any two adjacent discrete points. Alternatively, this implies that there is no scale in the problem that is smaller than the difference between the values the function takes at any two discrete points. In conventional metals this is usually the case: $\omega_c\ll \mu$ is the smallest energy scale in the problem, so summation over the Landau levels can be well approximated by Eq.~(\ref{eulermac}). Now, consider a situation where $E(\mb{k})$ has a feature at $\mb{k}=\tilde{\mb{k}}$ with energy $E(\tilde{\mb{k}})=\tilde{E}$ such that the slope changes sharply on a scale $\zeta$, and assume $\tilde{E}<\mu$. A magnetic field leads to Landau levels according to the semiclassical quantization formula, $S(E_n)l_B^2=2\pi(n+\gamma)$, where $S(E)$ is the area of an orbit at energy $E$ in $k$-space, $l_B=1/\sqrt{eB}$ is the magnetic length, and $\gamma$ is the semiclassical phase. Because of the feature $S(E)$ will also change sharply at $\tilde{E}$, and this will be reflected in the Landau level spectrum $E_n$: the slope of $E$ vs. $n$ will change abruptly leading to a change in level spacing from $\omega_{c1}$ to $\omega_{c2}$. Quantum oscillations are expected as long as $\zeta\ll\mathrm{max}\{\omega_{c1},\omega_{c2}\}$. Assume, for simplicity, $\zeta\ll\omega_{c1,2}$, i.e., $\zeta$ is the smallest energy scale in the problem. The Euler-MacLaurin formula cannot be applied directly anymore. However, note that only the Landau levels just below and above $\tilde{E}$ are effectively affected by $\zeta$. Hence, we first separate out these two Landau levels from the summation:
\beq
\frac{\Omega}{D}=\sum_0^{\tilde{N}-1}[E_n-\mu]+[E_{\tilde{N}}-\mu]+[E_{\tilde{N}+1}-\mu]+\sum_{\tilde{N}+2}^{N}[E_n-\mu].
\label{split}
\eeq
Here $\tilde{N}$ corresponds to the Landau level just below $\tilde{E}$ and $N$ corresponds to the Landau level just below $\mu$.
With this separation, we can now perform the sums using Eq.~(\ref{eulermac}). This reduces Eq.~(\ref{split}) to 
\begin{widetext}
\begin{eqnarray}
\frac{\Omega}{D}&\approx&\left[\int_{0}^{N}[E(n)-\mu]dn+\frac{1}{2}[\{E(N)-\mu\}+\{E(0)-\mu\}]+\frac{1}{12}[E'(N)-E'(0)]\right]\nonumber\\
&+&\left[
\int_{\tilde{N}+2}^{\tilde{N}-1}E(n)dn+\frac{1}{2}[E(\tilde{N}-1)+E(\tilde{N}+2)]+\frac{1}{12}[E'(\tilde{N}-1)-E'(\tilde{N}+2)]+E(\tilde{N})
+E(\tilde{N}+1)\right].
\label{omegaoscgen}
\end{eqnarray}
\end{widetext}
The first term in square brackets is the expression for the conventional case without any extra scale in the problem. It can be easily shown to give rise to three different contributions to $\Omega$: a part that does not depend on the field, a part that smoothly changes with the field---this gives rise to Landau diamagnetism in the susceptibility---and a part that oscillates with the inverse of the field---see Ref.\cite{sho} for a derivation. The second term in square brackets represents the correction due to the additional scale. This term is responsible for a new set of oscillations as anticipated earlier. Henceforth, we will refer to these oscillations as unconventional and those arising from the Fermi level as conventional. We emphasize that Eq.~(\ref{omegaoscgen}) is general and model-independent; it applies to any Landau level spectrum that satisfies the condition stated above, irrespective of its origin. Explicit calculations using the expression in Eq.~(\ref{omegaoscgen}) for the models considered in the next section are presented in the Supplementary Materials \cite{supp}.

\begin{figure*}
\centering
\subfigure[]{\includegraphics[width=.35\textwidth]{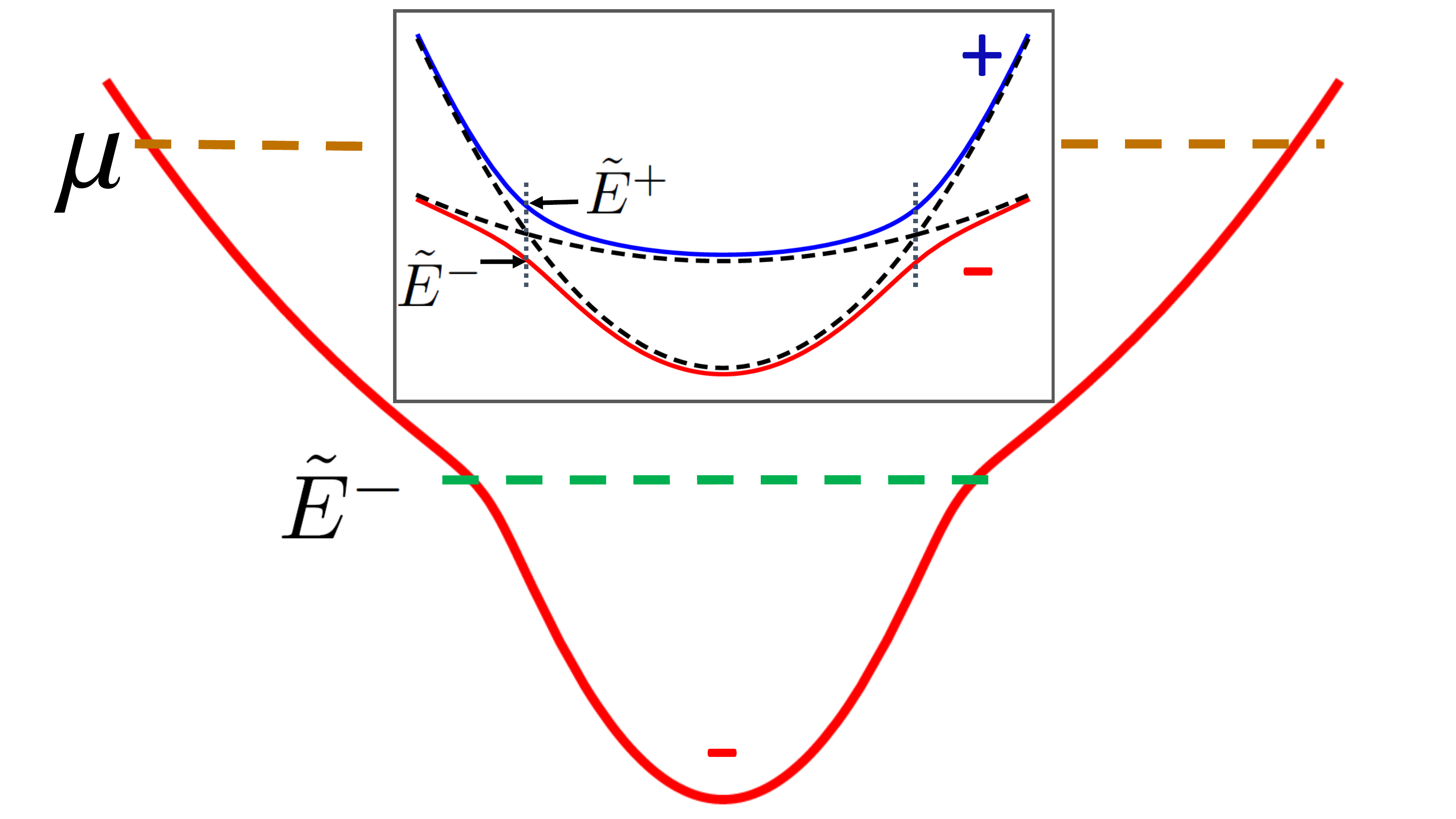}
\label{fig11}}
\quad
\subfigure[]{\includegraphics[width=.29\textwidth]{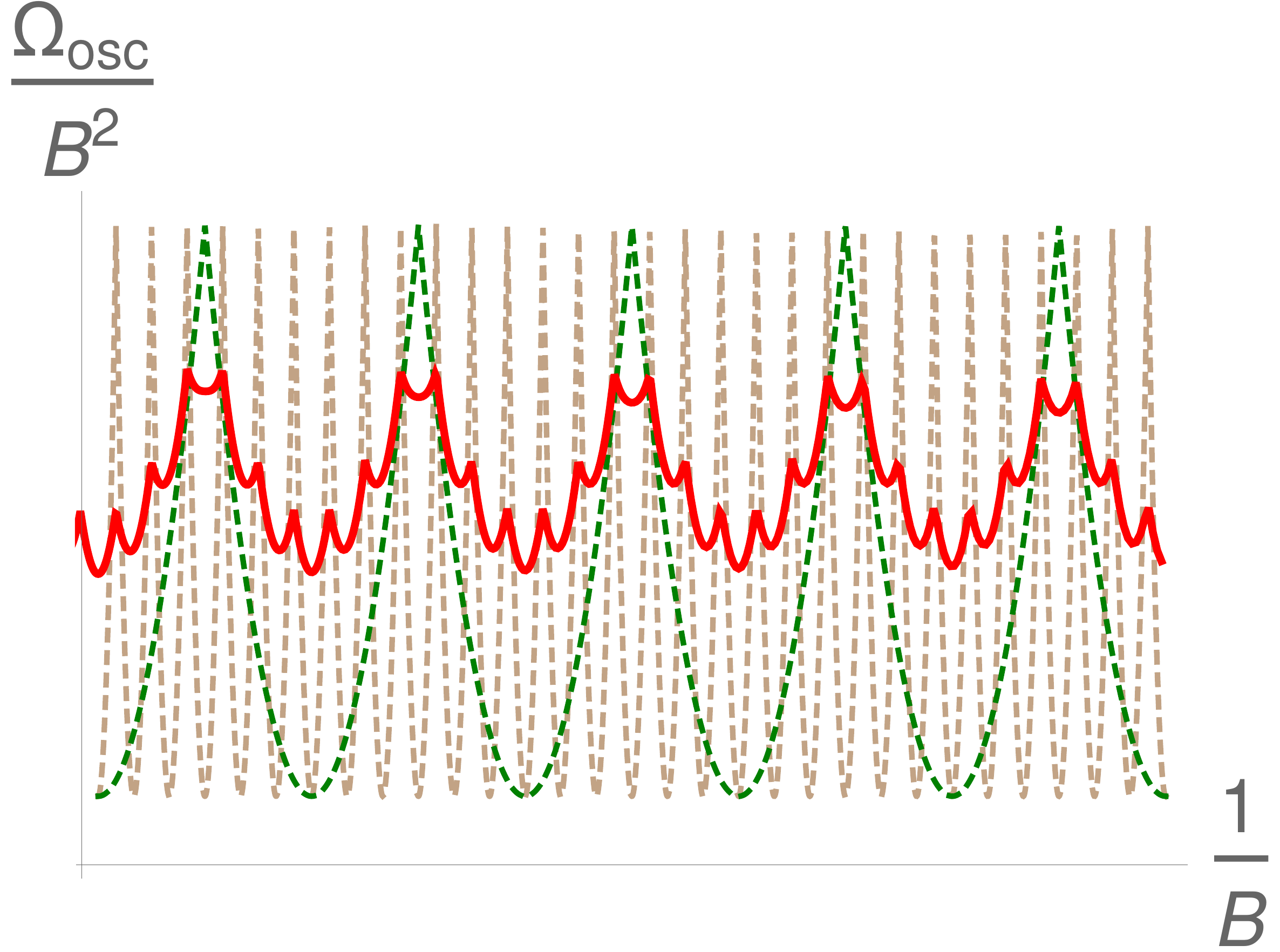}
\label{fig12}}
\quad
\subfigure[]{\includegraphics[width=.29\textwidth]{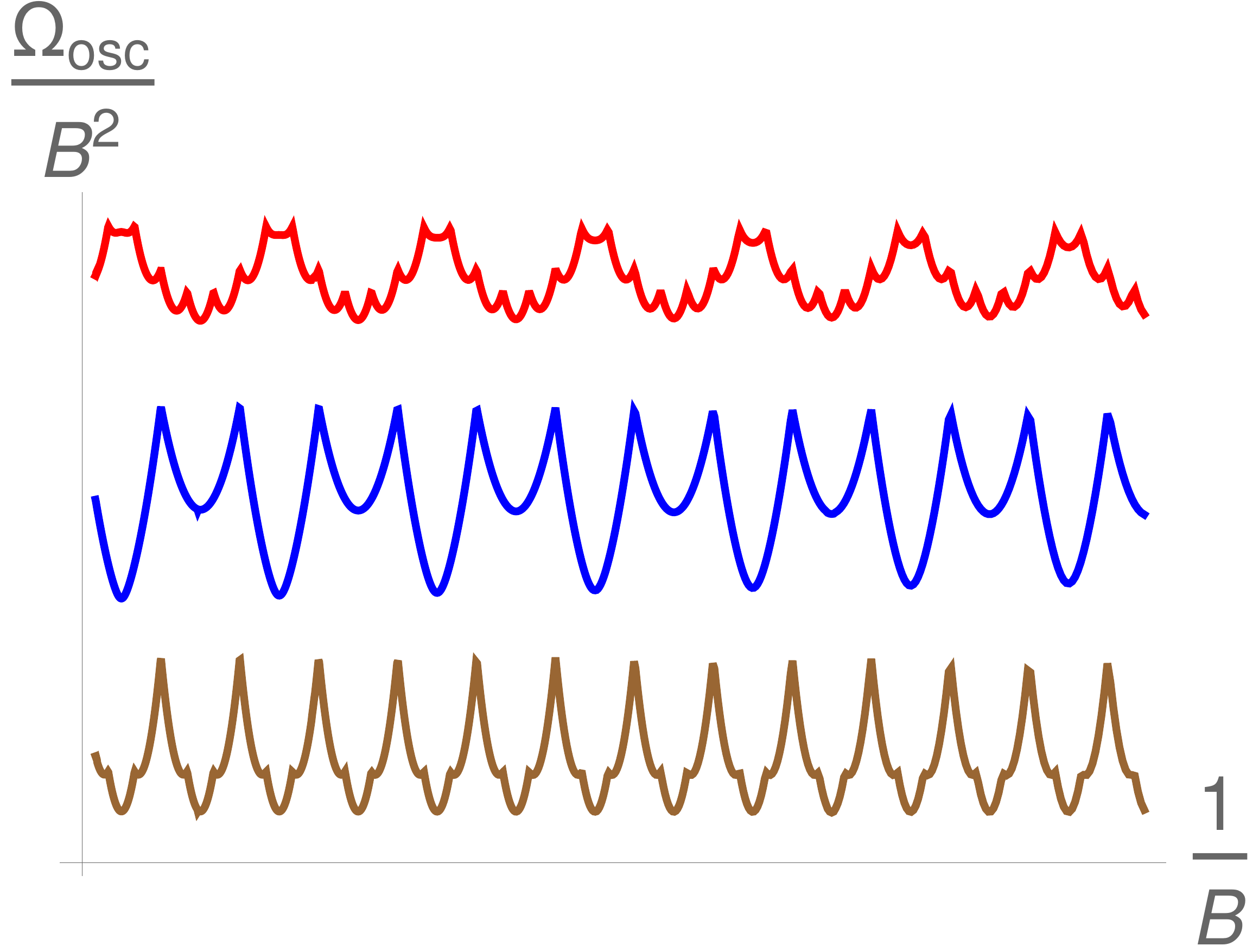}
\label{fig13}}
\caption{(a) Inset: Schematic band structure for the Hamiltonian in (\ref{ham}) with $\ve_1$ and $\ve_2$ having curvatures of same sign. Dotted curves show bands before hybridization. On hybridization, due to avoided crossing, the states at the intersection point are pushed up and down to $\tilde{E}^+$ and $\tilde{E}^-$. Main figure: One of the hybridized bands marked `-'. At $\tilde{E}^-$ the band slope changes abruptly and $\mu$ is the Fermi level. (b) Numerically calculated quantum oscillations for the $-$ band (red, solid). We chose $\ve_{1,2}=k^2/2m_{1,2}$, with $m_2/m_1=5$ and $\zeta/\Delta=.005$. Two frequencies are observed within a single band, one from $\mu$ and one from $\tilde{E}^-$. This is confirmed by plotting the oscillations expected from orbits at $\mu$ (brown, dotted) and $\tilde{E}^-$ (green, dotted). (c) Numerically calculated quantum oscillations for `-' band  (red, top), `+' band (blue, middle), and total (brown, bottom). The oscillations arising from $\tilde{E}^{-}$ and $\tilde{E}^+$ in the two bands are out of phase. Hence, they cancel out and do not show up in the total.}
\label{fig1}
\end{figure*}

\section{Models}

\begin{figure*}
\begin{center}
\includegraphics[scale=0.52,trim={0 8cm 0 0},clip]{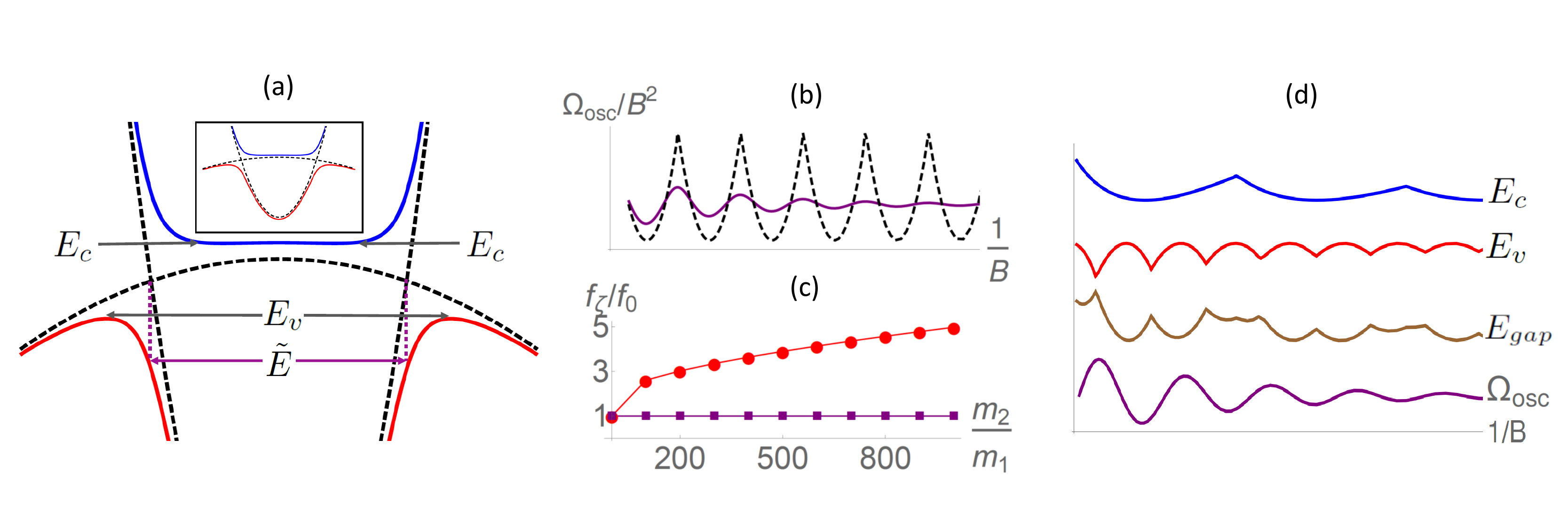}
\caption{(a) Inset: Same as in Fig.~\ref{fig1}(a) but with $\ve_{1}$ and $\ve_2$ having curvatures of opposite sign. This results in a gap. Main figure: When the Fermi level $\mu$ is in the gap, quantum oscillations arise from $\tilde{E}$ which is inside the valence band and not from the edge $E_v$. (b) Numerically calculated quantum oscillations for a lattice model simulating the band structure in (a): two square lattices with hopping parameters $t_1$ and $t_2$ ($t_2/t_1=-0.01$) intersecting at $1/8$ th filling were coupled with a hybridizing parameter ($\zeta/t_1=0.05$). $\mu$ is in the gap such that it would have passed through the band intersection before hybridization. The frequency of oscillations is same before and after hybridization. (c) Ratio of frequencies after and before hybridization, $f_{\zeta}/f_0$, as a function of the ratio of the masses of the hybridizing bands, $m_2/m_1=|t_1/t_2|$. The purple (square) curve is the numerically calculated curve while the red (circle) curve shows how frequency would have changed if oscillations originated at $E_v$ since area at the band edge increases with increase in $m_2/m_1$. The frequencies were extracted by finding the difference in $1/B$ values at which two consecutive maxima occur. (d) Numerically calculated oscillations for the highest VB energy level $E_v$, the lowest CB energy level $E_c$, the gap $E_{gap}=E_c-E_v$, and oscillating part of the grand canonical potential $\Omega_{osc}$. All curves in (c) and (d) were calculated based on the same lattice model used in (b).}
\label{fig2}
\end{center}
\end{figure*}

We now construct a simple model to demonstrate these unconventional oscillations. Consider two overlapping bands $\ve_1({\mb{k}})$ and $\ve_2({\mb{k}})$ with different masses 
hybridized by some parameter $\zeta$. In the band space, a general form of the Hamiltonian can be written as 
\beq
H_{\mb{k}}=
\begin{pmatrix}
\ve_1({\mb{k}})-\Delta&\zeta\\
\zeta&\ve_2({\mb{k}})
\end{pmatrix},
\label{ham}
\eeq
where $\Delta$ determines the overlap between the bands before hybridization.
The energy spectrum is given by
$E^{\pm}(\mb{k})=\frac{1}{2}\left[\ve_1({\mb{k}})+\ve_2({\mb{k}})-\Delta\pm\sqrt{\{\ve_2({\mb{k}})-\ve_1({\mb{k}})+\Delta\}^2+4\zeta^2}\right]$. Let $\mb{k}=\tilde{\mb{k}}$ denote the intersection between the two bands before hybridization. When the bands hybridize, due to avoided crossing, the slopes of the two resulting bands at $\tilde{\mb{k}}$ change sharply over a scale $\zeta$ at energies $\tilde{E}^{\pm}=E^{\pm}(\tilde{\mb{k}})$ with $\tilde{E}^+-\tilde{E}^-=2\zeta$ [Fig.~\ref{fig1}(a)]. This feature is also present in a magnetic field when Landau levels are formed:  $E^{\pm}(n)=\frac{1}{2}\left[\ve_1(n)+\ve_2(n)-\Delta\pm\sqrt{\{\ve_2(n)-\ve_1(n)+\Delta\}^2+4\zeta^2}\right]$, 
where $\varepsilon_{1,2}(n)$ denotes the Landau levels corresponding to $\ve_{1,2}({\mb{k}})$ with typical level spacing $\omega_c$. Assume $\zeta < \omega_c$ so that the feature is sharp. Such a system, therefore, satisifes the required condition for unconventional oscillations. Depending on the relative sign of the curvatures of the two unhybridized bands $\varepsilon_1$ and $\varepsilon_2$, i.e., their masses, two situations can arise: when $\varepsilon_1$ and $\varepsilon_2$ have curvatures of the same sign, the system remains metallic [Fig. \ref{fig1}(a)], and when $\varepsilon_1$ and $\varepsilon_2$ have curvatures of opposite sign, the system becomes an insulator with the opening of a gap [Fig. \ref{fig2}(a)]. 

We first consider the metallic regime [Fig. \ref{fig1}(a)]. Let $\mu$ be far above $\tilde{E}^\pm$. Consider one of the hybridized bands, say, the lower ($-$) band. Conventional understanding dictates that a single band should show oscillations with a single frequency arising from $\mu$.  Yet, in Fig.~\ref{fig1}(b) obtained numerically, we observe that $\Omega$ oscillates with two frequencies, one arising from the Fermi level and another arising from $\tilde{E}^-$ which is in the Fermi sea, in accordance to the predictions of our theory. Thermodynamic quantities obtained by appropriately differentiating $\Omega$, such as the magnetization and the susceptibility, will inherit the same two frequencies. For the calculation, we assumed $\ve_{1,2}(\mb{k})=k^2/2m_{1,2}$; however, the results are valid for any general spectrum. In fact, an exact analytical expression for the oscillations can be derived from Eq.~(\ref{omegaoscgen}) for arbitrary $\ve_{1,2}(\mb{k})$---see Supplementary Materials \cite{supp}. 

Although Fig.~\ref{fig1}(b) provides a clear confirmation of the predicted effect, an additional symmetry in this case renders the effect unobservable. Note that, similar to the `-' band, the `+' band also gives rise to it own set of oscillations arising from $\tilde{E}^+$ and $\mu$. As shown in Fig. \ref{fig1}(c), the unconventional oscillations arising in the two bands are exactly opposite in phase and cancel each other out. This happens because the features in the two bands are complementary: the sense in which the slope changes in one band is opposite to that in the other one. To prevent such cancellation, the additional symmetry has to be broken. One way to ensure this is to have $\tilde{E}^+>\mu>\tilde{E}^-$. Unfortunately, since $\zeta$ is small, in practice this will render the conventional and unconventional oscillations barely distinguishable. 

The above shortcoming, however, can be easily overcome if we consider the insulating regime of the same model (\ref{ham}). To this effect, we let $\ve_1(\mb{k})$ and $\ve_2(\mb{k})$ have curvatures of opposite sign. On hybridization, the system now becomes gapped [Fig.~\ref{fig2}(a)]. In this regime, if $\tilde{E}^+>\mu>\tilde{E}^-$ such that the Fermi level is in the gap, only the lower band can contribute to oscillations which cannot be  compensated by the upper band anymore since it is unoccupied. Further, because $\mu$ lies in the gap, no conventional oscillations can arise either; therefore, any oscillations in this model will be purely of the unconventional type. As in the previous case, the required feature due to hybridization is centered around the momentum $\tilde{\mb{k}}$ where the bands intersected before hybridization. One, therefore, expects unconventional oscillations arising from $\tilde{E}=E(\tilde{\mb{k}})$ as before (the `-' superscript is omitted for brevity since only one band contributes). It is easy to see that, as long as $\ve_1$ and $\ve_2$ are dissimilar, i.e., there is no particle-hole symmetry, $\tilde{E}<E_v$, where $E_v$ is the maximum energy of the VB---see Fig.~\ref{fig2}(a). Thus, the unconventional oscillations indeed originate from inside the Fermi sea. The frequency of oscillations is determined by the area of the orbit at $\tilde{E}$ which is same as the area of the orbit at the band intersection before hybridization. This implies that   the frequency of oscillations before hybridization when the Fermi level passes through the band intersection will be equal to the frequency of oscillations after hybridization when the Fermi level is in the gap. In Fig.~\ref{fig2}(b) we verify this numerically by calculating the oscillating part of the grand canonical potential exactly on a lattice model that mimics our starting Hamiltonian in (\ref{ham}). Thus, it can be seen that the origin of the unconventional oscillations in both the metallic and the insulating regime is the same; however, the latter has the advantage that it provides a realistic scenario where the unconventional oscillations can be experimentally measured. For such measurements one needs to ensure that $\zeta < \omega_c$, since the oscillations are inappreciable otherwise. As an estimate, using a mass of 0.01 time the bare electronic mass and fields $\sim 10$ T yields $\zeta < 100$ meV. The relation between $\zeta$ and the actual band gap depends on whether system is particle-hole symmetric or not: in the former case the band gap is equal to $2\zeta$ while in the latter case, in the limit of strong particle-hole asymmetry, the band gap is $\sim\zeta^2/\Delta$. Thus, narrow gapped materials are required. Note that, although the model (\ref{ham}) is written with simple bands in mind, the arguments above are much more general and can be extended to other systems that may not share the same low-energy model but nevertheless support similar gapped spectrum, such as gapped nodal-line semimetals \cite{fu,bia}, heavy Fermionic systems where itinerant electrons are hybridized with localized electrons \cite{dze}, materials at the onset of charge/spin density wave order where zone folding leads to hybridization between bands at different points in the parent Brillouin zone \cite{car,seb}, etc.

Recently, inspired by the experiment on SmB$_6$ in Ref.~\cite{tan}, special cases of the model above leading to a gap have been used to address the possibility of quantum oscillations in insulators \cite{kno,zha,pal}. In that context a few comments are in order.  Ref.~\cite{kno} considered the case where one of the hybridizing bands is flat, and found that the oscillations survive after hybridization. However, it is not obvious from where these oscillations arise in the absence of a Fermi surface.
This was addressed thereafter in Ref. \cite{zha} where it was argued that the oscillations originate from the band edge: the Landau levels periodically approach the edge resulting in oscillations in the gap causing quantum oscillations. Our theory not only provides an understanding for the origin of the oscillations reported in Ref. \cite{kno}, it also shows that the argument in Ref. \cite{zha} is, in general, not correct: the VB edge $E_v$ plays no role in oscillations, the latter are determined purely by $\tilde{E}$. Only in the case of a particle-hole symmetric model, i.e., when $\ve_1(\mb{k})=-\ve_2(\mb{k})$ in Eq.~(\ref{ham}), $\tilde{E}=E_v$, but this is merely a coincidence. As soon as particle-hole symmetry is broken, oscillations arise from inside the band \cite{com1}. One can argue that since $E_v-\tilde{E}\sim\mathcal{O}(\zeta)$, this difference in the origin leads to perturbatively small effects on oscillations and are unimportant. This is, however, not true. The area at the edge depends on both $\zeta$ and $m_2/m_1$, the ratio of the band masses controlling the particle-hole asymmetry. Choosing $m_2/m_1\gg 1$ offsets the smallness of $\zeta$ leading to an edge whose area is \emph{non-perturbatively} different from that at $\tilde{E}$. Therefore, if oscillations originated from the edge of the band, the frequency after hybridization should change drastically in a strongly particle-hole asymmetric system. In Fig.~\ref{fig2}(c) we confirm numerically for the same lattice model employed in Fig.~\ref{fig2}(b) that this is not the case: the frequency remains unchanged after hybridization irrespective of the particle-hole asymmetry. An even clearer picture emerges if we plot the oscillations of the gap directly. Since the area at the VB edge increases while that in the conduction band (CB) decreases [Fig.~\ref{fig2}(a)], we expect the highest energy level in the VB to oscillate with a higher frequency and the lowest energy level in the CB to oscillate with a lower frequency. This implies that the energy gap, given by their difference, will oscillate with a pattern comprising two frequencies. This is indeed what we observe in Fig.~\ref{fig2}(d) obtained from numerical calculations on the lattice as before. The curves show no resemblance to the oscillations in grand canonical potential. Thus, as far as oscillations at zero temperature are concerned, the gap itself has no significance---it is simply a by-product of the hybridization procedure. The same principle applies for the metallic case in Fig.~\ref{fig1}(a) and the insulating case in Fig.~\ref{fig2}(a). Oscillations always arise from $\tilde{E}$, independent of how the band behaves away from $\tilde{E}$---whether it is monotonic (metal) or non-monotonic (insulator)---which depends on band parameters. Also, our theory facilitates the determination of the frequency of oscillations  in hybridized bands, along with the origin, solely based on the given spectrum without requiring any knowledge of the bands prior to hybridization.  Conceptually, this is more agreeable, although, in practice, the argument could very well be used in reverse to advantage.

In addition to the examples discussed above, another possible scenario where quantum oscillations can arise from the Fermi sea without getting cancelled is when the quasiparticle dispersion features kinks arising from many-body effects. Usually the kinks result from the coupling of electrons with bosonic degrees of freedom, such as phonons \cite{lan,she} and spin excitations \cite{he,hwa}, although they can also arise from electron-electron interactions \cite{byc}. Irrespective of their origin, such features manifest themselves as a sharp change in the band slope away from the Fermi level. They have been reported both experimentally and theoretically in a variety of systems such as graphene, high T$_c$ superconductors, etc.\cite{lan,she,he,hwa,maz,kam}. In such cases, if the feature is sufficiently strong and persists in the presence of a magnetic field, the situation is no different from our example in Fig.~\ref{fig1}, except that now the oscillations from this feature are not compensated by those arising from a second band. Indeed, quantum oscillations may provide a useful tool to experimentally probe such features. 

\section{Discussion}

It is important to note that at zero temperature oscillations arising from the Fermi sea can only appear in thermodynamic quantities, i.e., quantities which are derived from the grand canonical potential, such as the magnetization and the susceptibility. Quantities which explicitly depend on the density of states at the Fermi level, such as the resistivity, cannot show such unconventional quantum oscillations.

Also, we have so far focused on the case of zero temperature; before concluding we provide a few remarks on the effect of temperature on these unconventional oscillations. While the oscillations at zero temperature can arise from within the Fermi sea, the effect of temperature is governed strictly by the Fermi level via the Fermi-Dirac function. Temperature reduces the amplitude of conventional oscillations due to dephasing. However, in the case of oscillations from Fermi sea, as long as $T\ll\mu-\tilde{E}$, it is obvious that temperature will have no effect. Indeed, if $\omega_c\ll\mu-\tilde{E}$, with increase in temperature, oscillations arising from the Fermi level will completely die leaving behind the unconventional oscillations unchanged. On the other hand, when $\omega_c>T\gtrsim\mu-\tilde{E}$ new effects can arise. In the case of gapped systems discussed above, where oscillations are only of the unconventional type, Refs. \cite{kno}, \cite{zha}, and \cite{pal} have already reported several nontrivial features in the temperature dependence in this regime. In cases where both conventional and unconventional quantum oscillations coexist, such as in metallic systems with kinks, one can expect a further new set of features in the temperature dependence in this regime hitherto unexplored.

Further, it should be noted that our results apply equally to both 2D as well 3D systems. In the 3D case, as in the case of metals, the oscillations will be governed by the extremal orbit perpendicular to the field. As long as the spectrum describing the orbit possesses the required feature where the band slope changes sharply, unconventional oscillations described above will arise.

\section{Concluding remarks}

In conclusion, we have shown that quantum oscillations can arise from inside the Fermi sea, in contrast to the conventional understanding that such oscillations always arise from the Fermi surface. These unconventional oscillations occur when the band slope changes abruptly on a scale that is smaller than the typical Landau level spacing in its vicinity, i.e., when there are kink-like features in the band. Such features can arise in insulators resulting from the hybridization of two overlapping bands with opposite curvature or from many-body effects. In particular, in strongly particle-hole asymmetric insulators, models of which have been recently used in the context of the topological Kondo insulator SmB$_6$, we have shown that the oscillations arise from inside the filled band and are not related to the gap. Our theory not only establishes a new paradigm for the phenomenon of quantum oscillations, it also provides a possible new way to experimentally explore certain features inside the Fermi sea.

\begin{acknowledgements}
I am grateful to F. Pi{\'e}chon and J-N. Fuchs for valuable discussions and for reading the manuscript, and to M. Goerbig for helpful suggestions. This work has been supported by the French program ANR DIRACFORMAG (ANR-14-CE32-0003) and LabEx PALM Investissement d'Avenir (ANR-10-LABX-0039-PALM).
\end{acknowledgements}

\begin{widetext}

\section{Supplementary Materials}

Consider a band $E(\mb{k})$ that has a feature at $\mb{k}=\tilde{\mb{k}}$ with energy $E(\tilde{\mb{k}})=\tilde{E}$ such that the slope changes sharply on a scale $\zeta$. In the presence of a magnetic field the energy spectrum becomes quantized. Because of the feature, the typical level spacing will change as well as one crosses $\tilde{E}$, say, from $\omega_{c1}$ to $\omega_{c2}$. Quantum oscillations are expected to arise from such a feature as long as $\zeta\ll\mathrm{max}\{\omega_{c1},\omega_{c2}\}$. In the present calculation, for simplicity, we assume $\zeta\ll\omega_{c1,2}$, i.e., the feature is sharp and $\zeta$ is the smallest energy scale in the problem. As shown in the main text, the grand potential $\Omega/D=\sum_{E_n\le\mu}[E_n-\mu]$, $D\propto B$ being the degeneracy of each Landau level (LL), can be written as
\begin{eqnarray}
\frac{\Omega}{D}&\approx&\left[\int_{0}^{N}[E(n)-\mu]dn+\frac{1}{2}[\{E(N)-\mu\}+\{E(0)-\mu\}]+\frac{1}{12}[E'(N)-E'(0)]\right]\nonumber\\
&+&\left[
\int_{\tilde{N}+2}^{\tilde{N}-1}E(n)dn+\frac{1}{2}[E(\tilde{N}-1)+E(\tilde{N}+2)]+\frac{1}{12}[E'(\tilde{N}-1)-E'(\tilde{N}+2)]+E(\tilde{N})
+E(\tilde{N}+1)\right]\nonumber\\
&=&T_1+T_2.
\label{omegaoscgen_supp}
\end{eqnarray}
Here $\tilde{N}$ corresponds to the LL just below $\tilde{E}$ and $N$ corresponds to the LL just below $\mu$. The first term in square brackets, $T_1$, is the expression for the conventional case without any extra scale in the problem.
The second term in square brackets, $T_2$, represents the correction due to the additional scale. This term is responsible for a new set of unconventional oscillations from Fermi sea. In the following we compute explicit analytical expressions for  these unconventional oscillations for a simple model.

\begin{figure}
\centering
\subfigure[]{\includegraphics[width=.31\textwidth]{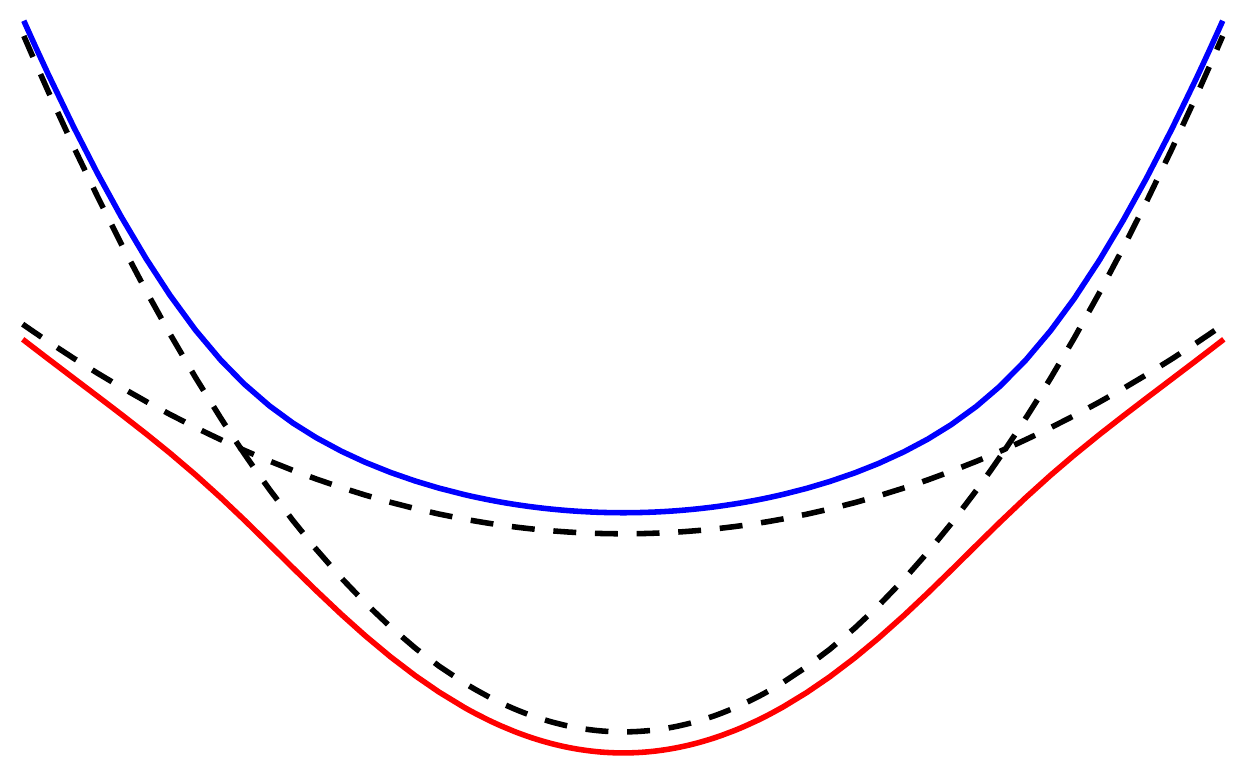}
\label{fig11}}
\quad
\hspace{2cm}
\subfigure[]{\includegraphics[width=.30\textwidth]{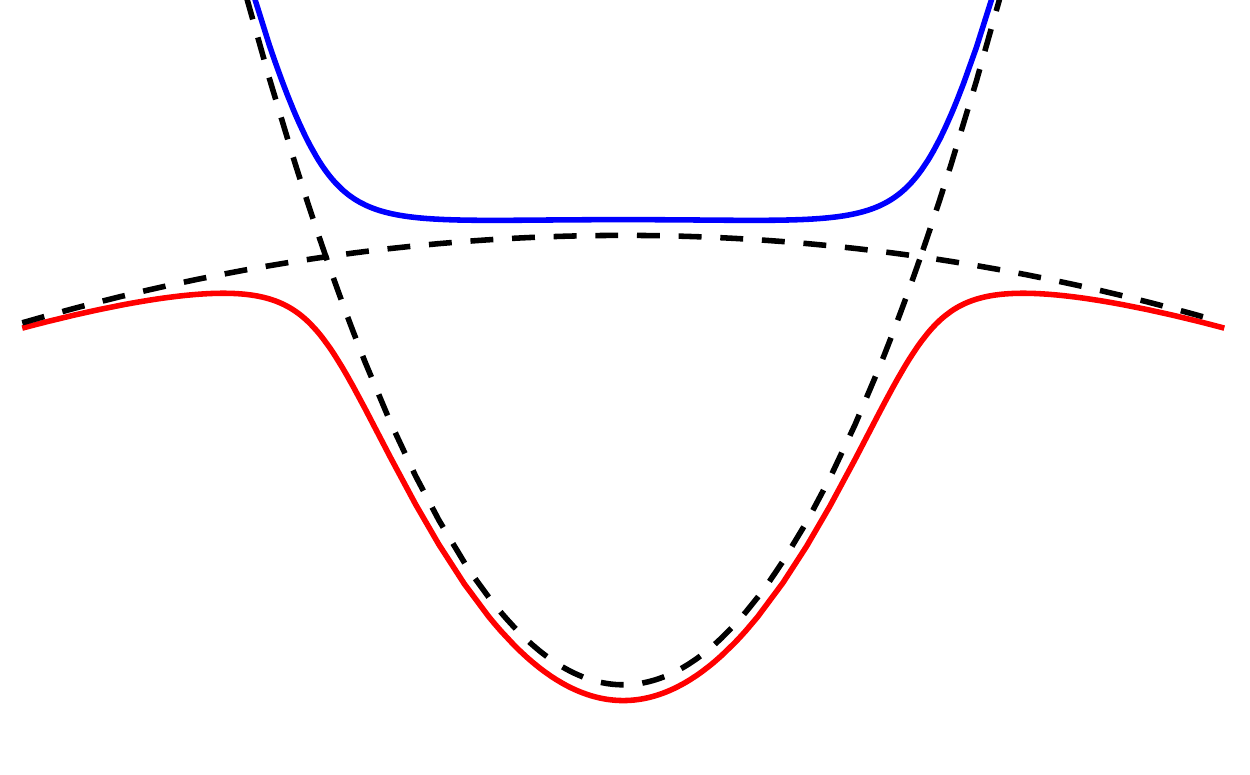}
\label{fig12}}
\caption{Schematic band diagrams for the Hamiltonian in (\ref{ham}) for (a) $m_1/m_2>0$ and (b) $m_1/m_2<0$, where $m_{1,2}$ are the masses of the two bands at the point of intersection before  hybridization. In the former case, the system remains metallic, while in the latter case it becomes insulating due to the appearance of a gap.}
\label{fig1_supp}
\end{figure} 

The model considered in the main text is that of two overlapping bands $\ve_1({\mb{k}})$ and $\ve_2({\mb{k}})$ with masses $m_1$ and $m_2$, calculated at the intersection point and not equal to each other. Hybridization between the two bands results in avoided crossing. In the band space, a general form of the Hamiltonian can be written as (we set $\hbar=k_B=1$)
\beq
H_{\mb{k}}=
\begin{pmatrix}
\ve_1({\mb{k}})-\Delta&\zeta\\
\zeta&\ve_2({\mb{k}})
\end{pmatrix},
\label{ham}
\eeq
where $\Delta$ determines the overlap between the bands before hybridization, and $\zeta$ is the hybridizing parameter. 
The energy spectrum is given by
$E^{\pm}(\mb{k})=\frac{1}{2}\left[\ve_1({\mb{k}})+\ve_2({\mb{k}})-\Delta\pm\sqrt{\{\ve_2({\mb{k}})-\ve_1({\mb{k}})+\Delta\}^2+4\zeta^2}\right]$. The resulting band structure is metallic for $m_1/m_2>0$ and insulating for $m_1/m_2\le 0$---see Fig.~\ref{fig1_supp}. Let $\mb{k}=\tilde{\mb{k}}$ denote the intersection between the two bands in the absence of hybridization. After hybridization, due to avoided crossing, the slopes of the bands at $\tilde{E}^{\pm}=E^{\pm}(\tilde{\mb{k}})$ change sharply over a scale $\zeta$. This feature is also present in the presence of magnetic field when LLs are formed:  
\beq
E^{\pm}(n)=\frac{1}{2}\left[\ve_1(n)+\ve_2(n)-\Delta\pm\sqrt{\{\ve_2(n)-\ve_1(n)+\Delta\}^2+4\zeta^2}\right], 
\eeq
where $\varepsilon_{1,2n}$ denotes the LLs corresponding to $\ve_{1,2}({\mb{k}})$. We wish to evaluate Eq.~(\ref{omegaoscgen_supp}) for this model. We will calculate only for the -- band, the calculation for the other band (+) is identical. Henceforth, we drop the -- superscript for brevity.

Before deriving the unconventional oscillations, let us first show how the conventional ones arise. To do so we follow the method given in Ref.~\cite{shoenberg}. The same method will be adopted to derive the unconventional case.
The general relation between $E$ and $n$ is given by the semiclassical quantization condition,
\beq
S(E)l_B^2=2\pi(n+\gamma),
\label{quantcon}
\eeq
where $S(E)$ is the area of an orbit in $k$-space as a function of $E$, $l_B=1/\sqrt{eB}$ is the magnetic length, $n$ is the LL index, and $\gamma$ is the semiclassical phase. To proceed further, it is useful to define a variable $x$ in place of $n+\gamma$ and rewrite the quantization condition as
\beq
S(E)l_B^2=2\pi x.
\eeq
Let $X$ be the value $x$ takes at the Fermi level, i.e., $S(\mu)l_B^2=2\pi X$. With change in magnetic field, the LLs move, and each time a LL crosses $\mu$, $X$ changes by one. With this in mind define $\delta=X-(N+\gamma)$ so that $0\le\delta <1$. Inserting this in Eq.~(\ref{omegaoscgen_supp}), we have
\beq
T_1\approx\int_{\gamma}^{X-\delta}[E(x)-\mu]dx+\frac{1}{2}[\{E(X-\delta)-\mu\}+\{E(\gamma)-\mu\}]+\frac{1}{12}[E'(X-\delta)-E'(\gamma)].
\label{omegasumX}
\eeq
Expanding around $X$, after some algebra, it can be reduced to
\beq
T_1\approx\int_{0}^{X}[E(x)-\mu]dx+\left(\frac{1}{2}-\gamma\right)\left[E(0)-\mu\right]+\omega_c(0)\left(-\frac{\gamma^2}{2}+\frac{\gamma}{2}-\frac{1}{12}\right)+\frac{1}{2}\omega_c(\mu)\left(\delta^2-\delta+\frac{1}{6}\right),
\label{omegasumX2}
\eeq
where $\omega_c=\partial E/\partial n$. The last term depending on $\delta$ is responsible for oscillations \cite{comment}. Thus, we have for the conventional oscillations arising from the Fermi level,
\beq
\frac{\Omega_{osc}^{\mu}}{D\omega_c}=\frac{1}{2}\left(\delta^2-\delta+\frac{1}{6}\right).
\label{osccon}
\eeq
The above expression is valid for $0\le\delta< 1$ and it gives one period of oscillation. Fig.~\ref{fig2_supp}(a) shows a plot of Eq.~(\ref{osccon}). This pattern must be repeated to get complete oscillations.

\begin{figure}
\includegraphics[width=.99\textwidth]{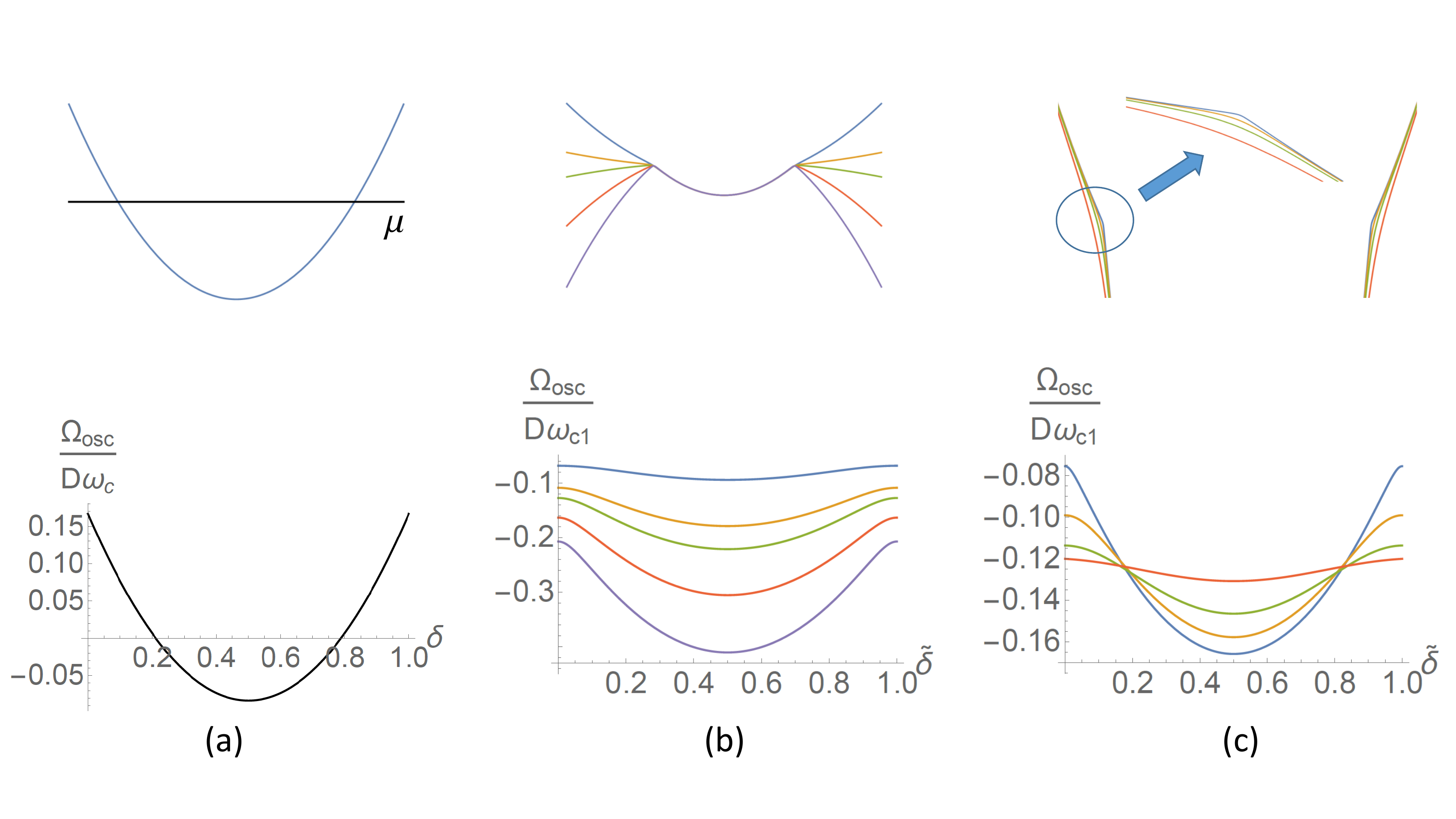}
\caption{(a) $\Omega_{osc}$ vs. $\delta$ for conventional oscillations arising from the Fermi level [Eq.~(\ref{osccon})]. Panel on top shows the band structure. (b) and (c) $\Omega_{osc}$ vs. $\tilde{\delta}$ for unconventional oscillations arising from the feature where the band slope changes suddenly within the Fermi sea [Eq.~(\ref{oscuncon})] for different values of the parameters $a$ and $b$. Here $a=\zeta/\omega_{c1}$ and $b=m_1/m_2$.  Panels on top show the  corresponding band structures for the different parameter values (each curve in the bottom panel corresponds to the band with the same color in the top panel). In (b) we have used $a=0.05$ and $b=0.5,0.1,-0.1,-0.5,-1.0$ (going downwards in the figure).When $b\le 0$, the bands are gapped, as opposed to $b>0$ when they are metallic. The figure shows that it is the feature that contributes to oscillations irrespective of whether the resulting band structure is that of a metal or an insulator. Also, it shows that the greater the change in slope at the feature, the bigger the amplitude of the oscillations. In (c) we have used $b=0.2$ and $a=0.01,0.05,0.1,0.2$ (going upwards in the figure). The figure demonstrates that the sharper the feature, the more pronounced the oscillations, i.e., the amplitude is larger.}
\label{fig2_supp}
\end{figure}

Let us now calculate the term $T_2$ in Eq.~(\ref{omegaoscgen_supp})---the term responsible for unconventional oscillations---in the same spirit. To that effect, define $\tilde{X}$ as the value $x$ takes at $\tilde{E}$, i.e., $S(\tilde{E})l_B^2=2\pi \tilde{X}$. With change in magnetic field, the LLs move, and each time a LL crosses $\tilde{E}$, $\tilde{X}$ changes by one. Define $\tilde{\delta}=\tilde{X}-(\tilde{N}+\gamma)$ so that $0\le\tilde{\delta}<1$. Inserting this in Eq.~(\ref{omegaoscgen_supp}), we have
\beq
T_2=\int_{\tilde{X}-\tilde{\delta}+2}^{\tilde{X}-\tilde{\delta}-1}E(x)dx+\frac{1}{2}[E(\tilde{X}-\tilde{\delta}-1)+E(\tilde{X}-\tilde{\delta}+2)]+\frac{1}{12}[E'(\tilde{X}-\tilde{\delta}-1)-E'(\tilde{X}-\tilde{\delta}+2)]+E(\tilde{X}-\tilde{\delta})
+E(\tilde{X}-\tilde{\delta}+1)
\label{oscsum2}
\eeq
Unlike before, we cannot expand near $\tilde{X}$ anymore, since $E(x)$ changes rapidly on the scale $\zeta$ near $\tilde{X}$. However, one can still expand the original unhybridized bands $\ve_{1,2}$ near $\tilde{X}$:
\beq
\ve_{1,2}(\tilde{X}+\tilde{\delta})\approx\ve_{1,2}(\tilde{X})+\omega_{c1,2}\tilde{\delta},
\eeq
where $\omega_{c1,2}=\partial\ve_{1,2}/\partial n$. This is justified as long as the effective masses of the unhybridized band do not change on the scale of $\zeta$, which by assumption is valid since $\zeta$ is the smallest energy scale in the problem. Thus we have
\beq
E(\tilde{X}+\tilde{\delta})\approx\frac{1}{2}\left[(\omega_{c1}+\omega_{c2})\tilde{\delta}-\sqrt{(\omega_{c1}-\omega_{c2})^2\tilde{\delta}^2+4\zeta^2}\right],
\label{edgespectrum}
\eeq
where we have used $\ve_{1}(X)=\Delta$ and $\ve_2(X)=0$. Note that the above expression is independent of the details of the spectrum of the original hybridizing bands: all information about the underlying spectrum is now contained in the parameters $\omega_{c1,2}$ which in turn depends on the effective masses $m_{1,2}$ computed at $\tilde{E}$.
Inserting Eq.~(\ref{edgespectrum}) into Eq.~(\ref{oscsum2}), we have the term for unconventional oscillations arising from inside the Fermi sea:
\beq
\frac{\Omega_{osc}^{\tilde{E}}}{D\omega_{c1}}=\frac{1}{24} (b -1)^2\left(\frac{\tilde{\delta} -2}{g_{\tilde{\delta}-2}}-\frac{\tilde{\delta} +1}{g_{\tilde{\delta}+1}}\right)+\frac{1}{4}\left[(1-\tilde{\delta}) g_{\tilde{\delta}-2}-2 g_{\tilde{\delta}-1}-2 g_{\tilde{\delta}}+\tilde{\delta}  g_{\tilde{\delta}+1}\right]+\frac{a^2}{b -1}\log \left[\frac{(b -1) (\tilde{\delta} +1)+g_{\tilde{\delta}+1}}{(b -1) (\tilde{\delta} -2)+g_{\tilde{\delta}-2}}\right],
\label{oscuncon}
\eeq
with
\beq
g_{\tilde{\delta}}=\sqrt{4 a^2+(b -1)^2 \tilde{\delta} ^2},
\eeq
and $a=\zeta/\omega_{c1}$ and $b=m_1/m_2$. Figs.~\ref{fig2_supp}(b) and (c) show plots for the unconventional oscillations for different values of the parameters $b$ and $a$. As before, the plots show one period of oscillation. One must repeat this to get the complete oscillation pattern. It can be seen that the amplitude of the oscillations depends both on the difference in the slopes of the two bands that gives rise to the feature, determined by $b$, as well as how fast the change of slope happens, i.e., the sharpness of the feature, determined by $a$: the bigger the change in slope and sharper the change, the bigger the amplitude of oscillations. Also, unlike before, the amplitude decreases as $1/B$ increases (since $\zeta/\omega_{c1}$ increases), thus the oscillations become damped as a function of the inverse field. Note that for $b>0$, the system stays a metal, but becomes an insulator for $b\le 0$ since a gap opens up. The above analysis clearly shows that the unconventional oscillations discussed here arise purely due to the feature inside the Fermi sea and has no bearing on whether the system is a metal or an insulator which, in the present context, is just a by-product of the hybridization and plays no role in oscillations at zero temperature.

\end{widetext}


\begin{thebibliography}{99}
\bibitem{supp} [URL will be inserted by publisher]
\bibitem{com1} Ref.~\cite{pal} also used the same argument as in Ref.~\cite{zha} that oscillations in gapped systems arise from the edge of the valence band. However, because the model used in Ref.~\cite{pal} is particle-hole symmetric, the argument there is not inconsistent with the theory in the present paper.
\end{thebibliography}
\begin{thebibliography}{99}
\bibitem{comment} Strictly speaking, this expression is only valid for parabolic spectrum where $\gamma=1/2$. For a general spectrum, $\gamma$ is energy dependent and hence leads to corrections in Eq.~(\ref{omegasumX2}). This modifies the part leading to Landau diamagnetism already in the leading order and is important (see Ref.~\cite{shoenberg}); however, the correction for the oscillating part is in the next-to-leading order. Therefore, as far as the oscillating part is concerned, the expression is valid for any general spectrum.
\end{thebibliography}
\end{document}